\documentclass[aps, pra, preprint, groupedaddress, amsfonts,
               amsmath, amssymb, showpacs, nofootinbib]{revtex4-1}
\usepackage{microtype}
\usepackage{graphicx}
\usepackage{amssymb}
\usepackage{mathtools}
\usepackage{epstopdf}
\usepackage[utf8]{inputenc}
\usepackage[T1]{fontenc}
\usepackage[usenames,dvipsnames]{xcolor}
\usepackage{hyperref}

\usepackage{color}

\begin{document}
\title{Classical-trajectory time-dependent mean-field theory for ion-molecule
collision problems}

\author{Alba Jorge}  
\email[]{albamaria.jorge@gmail.com}
\affiliation{Departamento de Qu\'imica, Universidad Aut\'onoma de Madrid, Cantoblanco, E-28049 Madrid, Spain}

\author{Marko Horbatsch}  
\email[]{marko@yorku.ca}
\affiliation{Department of Physics and Astronomy, York University, Toronto, Ontario, Canada M3J 1P3}

\author{Tom Kirchner}  
\email[]{tomk@yorku.ca}
\affiliation{Department of Physics and Astronomy, York University, Toronto, Ontario, Canada M3J 1P3}
\date{\today}
\begin{abstract}
A mean-field model to describe electron transfer 
processes in ion-molecule collisions at the
$\hbar =0$ level is presented and applied to collisions involving
water and ammonia molecules.
Multicenter model potentials account for the molecular
structure and geometry. They include charge screening parameters
which in the most advanced version of the model depend on the
instantaneous degree of ionization so that
dynamical screening effects are taken into account.
The work is implemented using the
classical-trajectory Monte Carlo method,
i.e., Hamilton's equations are solved for 
classical statistical ensembles that represent the
initially populated orbitals.
The time-evolved trajectories are sorted into
ionizing and electron capture events, and a multinomial
analysis of the ensuing single-particle probabilities
is employed to calculate differential and total cross
sections for processes that involve single- and
multiple-electron transitions.
Comparison is made with experimental data and 
some previously reported
calculations to shed light on the
capabilities and limitations of the approach.

\end{abstract}

\maketitle
\section{Introduction}
\label{intro}
Atoms and molecules are objects of the microscopic world. Hence, it seems
natural to use quantum mechanics to describe their structure and interactions.
What is natural is however not necessarily easy and perhaps not strictly
required, at least as long as typical actions are much larger than
Planck's constant.
Put in more quantitative terms, in the $\hbar \rightarrow 0$ limit
the quantum-mechanical
Schr\"odinger equation can be replaced by classical equations of motion, which
are easier to solve, and the time evolution of the quantum system be 
understood at the level of classical statistical mechanics~\cite{messiah}.

In the context of atomic collisions this insight forms the basis of the classical-trajectory
Monte Carlo (CTMC) method. It was introduced a long time ago~\cite{Abrines_1966} 
and has since been used
in many different variants for a multitude of scattering problems, starting from the prototypical one-electron
proton-hydrogen system, carrying over to multicharged-ion many-electron-atom systems, and
more recently also finding application in problems involving molecular targets. 
A very large number of studies
have been carried out and in many cases provided the first (and sometimes the only) 
theoretical results capable of describing complex experimental data~\cite{Ullrich_1997}.

It is beyond the scope of this chapter to provide a comprehensive overview of all these works. Interested readers
are referred to~\cite{Cocke_1991, Fremont_2021} for reviews and lists of references to the 
original literature. 
Our objective is more modest and
specific: We describe a recently proposed CTMC method that extends the standard approach by
incorporating dynamical screening effects in the
spirit of a time-dependent mean-field model, and show by way of a few illustrative examples what can
be learned from it when it is applied to the problem of single and multiple electron removal in
ion-molecule collisions. We study total and differential cross sections. 
Both are needed for an accurate understanding of, e.g., the problem of radiation damage of biological
tissue, which has been a major driver of contemporary ion-molecule collision studies.
Differential data for electron emission provide more
detailed information than total ionization cross sections 
but are harder to calculate on the basis of quantum-mechanical methods, especially
those which do not make use of perturbative arguments.
This makes CTMC calculations all the more important
and motivated us to pursue this work. 

Ion-molecule collision systems are many-particle problems. We address them in the frameworks
of the semiclassical impact parameter approximation and the independent electron model (IEM),
which also form the basis of most quantum-mechanical approaches to the problem at hand.
For us this means that the $\hbar \rightarrow 0$ limit is applied to a one-electron time-dependent
Schr\"odinger equation whose Hamiltonian includes a mean-field potential to account
for the electron-electron interactions in an effective manner
and
which has to be solved
for the molecular orbitals (MOs) which form the molecular ground state, i.e., the initial state
of the target before it is approached by the projectile ion.

The mean-field potential model used has a multicenter structure reflecting the molecular
geometry and takes time-dependent screening effects into account.
Our published work focuses on ion collisions with water (H$_2$O) 
molecules~\cite{alba_2019, alba_2020, hjl_2020, Bhogale_2022},
which are of paramount interest for the radiation damage problem mentioned above,
but more recently we have also looked at the proton-ammonia (NH$_3$) system~\cite{alba_2023},
and we discuss it alongside the ion-water problem in this chapter.

The layout is as follows. We begin in Sect.~\ref{sec:theory} 
with a short summary of the general theoretical assumptions and ideas
on which our model description is built. This is followed by somewhat
more detailed discussions of the (time-dependent)
mean-field potential models (in Sect.~\ref{sec:potential}) and the 
procedures used to extract the desired information, i.e., 
collisional cross sections for various electron removal processes, 
from the solutions of the equations of motion (in Sect.~\ref{sec:analysis}).
Results are presented and compared with experimental data and 
selected previous calculations
in Sect.~\ref{sec:results}, first for
H$_2$O target molecules (in Sects.~\ref{sec:water-dcs} and \ref{sec:water-tcs}) and then
for NH$_3$ (in Sect.~\ref{sec:ammonia}). The chapter ends with
a few concluding remarks in Sect.~\ref{sec:conclusions}.
Atomic units, characterized by $\hbar=m_e=e=4\pi\epsilon_0=1$, are used unless otherwise stated.

\section{Theory}
\label{sec:theory}
We are concerned with relatively high-impact-energy ion-molecule collisions
so that we can assume the projectile ion to travel on a straightline trajectory
and the target molecule to remain rigid during the (short) interaction time.
This is to say that we are using the 
semiclassical impact parameter approximation 
and keep the rotational and vibrational degrees of freedom of the target frozen.
To deal with the many-electron nature of the problem we apply the IEM, i.e., we
assume the Hamiltonian to be of single-particle form ($H \approx \sum_j h^{(j)}$)
so that the equations of
motion have one-electron character. Solving them for the MOs that form the
molecular ground state provides all accessible information on the electronic
transitions taking place during the collision. 
The single-electron Hamiltonian has the standard form
\begin{equation}
	h= \frac{{\bf p}^2}{2} + v_{\rm mod} + v_{\rm p} ,
\label{eq:hamilton}
\end{equation}
where $v_{\rm mod}$ denotes the multicenter target model potential and $v_{\rm p}$ 
the projectile potential. Both are specificied in Sect.~\ref{sec:potential}
and together define the (time-dependent) mean-field model used.

Applying the $\hbar\rightarrow 0$ limit amounts in practice
to solving Hamilton's equations for large sets
of initial conditions which represent discretized versions of
microcanonical ensembles of trajectories corresponding to the ionization energies of the MOs.
For H$_2$O, a more detailed discussion of how the initial distributions 
are generated (along with plots to show them)
is provided in~\cite{Illescas_2011}. An analogous procedure is used for NH$_3$.

The trajectories are time evolved using a small step size of $\Delta t=0.05$ a.u. 
until the projectile has reached a distance of about 500 a.u. to the target at the
final time $t=t_f$
after passing it at a specific impact parameter vector. 
For each trajectory we look at the energy of the
electron with respect to the projectile and the target in order to determine
its fate: If 
one of them is negative the electron is 
considered bound to that center, i.e., it is either captured by the projectile
or it remains bound to the target. If 
both energies are positive, the trajectory contributes to a transition to the
continuum (ionization). 

Calculations are carried out for many orientations of the target molecule
characterized by randomly chosen
sets of Euler angles, and orientation averages are calculated at the end in
order to allow comparisons with orientation-insensitive experimental data.
The models used to extract the differential and total cross sections
for the processes of interest are described
in Sect.~\ref{sec:analysis}.

\subsection{Mean-field potential models}
\label{sec:potential}
The ideas that go into setting up the 
(time-independent and time-dependent) mean-field
potentials are the same for water and ammonia.
Hence, we describe them for both target molecules in tandem.

\subsubsection{Static screening model}
\label{sec:static}
The multicenter target ($T$) potentials are approximated as 
sums of (time-independent) central potentials for each atom of the molecule:
\begin{eqnarray}
v_{\rm mod} &=& \sum_i v_{\rm X}(r_{{\rm X}_i}) \label{eq:pot1} \\
	    &=& v_{\rm O}(r_{\rm O}) + \sum_{i=1}^2 v_{\rm H}(r_{{\rm H}_i}) , \hskip20pt
	    (T={\rm H}_2{\rm O}) , \\
            &=& v_{\rm N}(r_{\rm N}) + \sum_{i=1}^3 v_{\rm H}(r_{{\rm H}_i}) , \hskip20pt
	    (T={\rm NH}_3) . 
\end{eqnarray}
In~(\ref{eq:pot1}), X labels the atoms (X=O,N,H) and
$r_{{\rm X}_i}$ are the distances from the
active electron to the $i=1,\ldots ,M$ nuclei of the molecule.
Note that H$_2$O is planar while NH$_3$ is umbrella shaped.
We use the following nuclear geometry parameters: 
The O-H bond lengths in H$_2$O are fixed at 1.8 a.u. and the angle between the
position vectors for the two protons is 105$^\circ$. 
The N-H bond lengths in NH$_3$ 
are 1.928 a.u., the 
azimuthal angles of the proton position vectors are 90$^\circ$, 210$^\circ$, and 330$^\circ$
and their polar angles are 108.9$^\circ$~\cite{Moccia_1964b}. 
A planar geometry would correspond to
right polar angles.
The central potentials in~(\ref{eq:pot1}) are assumed to be of the form
\begin{equation}
v_{\rm{X}}(r_{\rm{X}})=-\frac{Z_{\rm{X}}-N_{\rm{X}}}{r_{\rm{X}}}-\frac{N_{\rm{X}}}{r_{\rm{X}}}(1+\alpha_{\rm{X}}r_{\rm{X}})\exp(-2\alpha_{\rm{X}}r_{\rm{X}}) ,
\label{eq:pot2}
\end{equation}
where $Z_{\rm{X}} = 8,7,1$ are the nuclear charge numbers of X$=$O, N, H, while the other parameters 
have the values $N_{\rm{O}} = 7.185$, $\alpha_{\rm{O}}=1.602$, $N_{\rm{N}} = 6.2775$, 
$\alpha_{\rm{N}}=1.525$, $N_{\rm{H}} = 0.9075$, and $\alpha_{\rm{H}}=0.6170$. 
These values were found as follows:
At very large distances the total potential experienced by any one electron of a neutral atom or molecule behaves like $-1/r$. This condition results in the relations
$N_{\rm{O}}+2 N_{\rm{H}}=9=N_{\rm{N}}+3 N_{\rm{H}}$.
Using the first equality as a constraint, the parameters for H$_2$O were determined by
minimizing the differences of the valence MOs' energy eigenvalues found for  
the (quantum-mechanical) Hamiltonian $\frac{{\bf p}^2}{2} + v_{\rm mod} $ with respect to 
self-consistent field (SCF) results~\cite{Illescas_2011}.
For NH$_3$ we use the same parameters $N_{\rm{H}}$ and $\alpha_{\rm{H}}$, find
$N_{\rm{N}}=9-3 N_{\rm{H}}=6.2775$, and determine 
the remaining parameter $\alpha_{\rm{N}}$ such that
the energy eigenvalues of $\frac{{\bf p}^2}{2} + v_{\rm mod} $ are in reasonable 
agreement with the SCF results from~\cite{Moccia_1964b}. 

If dynamical screening is excluded, the projectile potential 
of a fully-stripped ion is purely Coulombic: $V_{\rm p} = -Z_{\rm p}/r_{\rm p}$
with $Z_{\rm p}$ and $r_{\rm p}$ being the charge number of the projectile and
the distance from it to the active electron.
We treat the partially-stripped Si$^{13+}$ ion for which we 
present results in Sect.~\ref{sec:water-dcs} in the same way, i.e., by a bare
Coulomb potential corresponding to the asymptotic charge.

\subsubsection{Dynamical screening model}
\label{sec:dynamic}
Consider a small-impact-parameter collision of a highly-charged projectile
ion. Around the distance of closest approach the projectile penetrates the molecule
and exerts a strong force on the electrons. The likely outcome is {\it multiple}
electron removal which increases the pull of the now partially ionized target on
the electrons.

This dynamical effect can be modelled in a simple fashion by replacing the static
screening charge parameters $N_{\rm X}$ in~(\ref{eq:pot2}) by time-dependent ones
in a way that ensures that the screening decreases with increasing degree of ionization $q$. 
The instantaneous degree of ionization corresponds to the (fractional) average number 
of removed electrons, the net electron removal $P^{\rm{rem}}_{\rm{net}}$.
Hence, we can feed information on the degree of ionization into the potential by
making the charge screening parameters 
dependent on $P^{\rm{rem}}_{\rm{net}}$. We use the simple ansatz
\begin{equation}
N_{\rm{X}}(P^{\rm{rem}}_{\rm{net}})=
\begin{cases}
	\begin{alignedat}{2}
	&N_{\rm{X}}^c\:\:\:\:\:\:\:\:\:\:&& P^{\rm{rem}}_{\rm{net}}\leq 1 ,\\
	&\frac{N_{\rm{X}}^c N_T}{N_T-1}\left(1-\frac{P^{\rm{rem}}_{\rm{net}}}{N_T}\right)\:\:\:\:\:\:\:\:\:\:&& 
	1\leq P^{\rm{rem}}_{\rm{net}}\leq N_T ,\\
	\end{alignedat}
\end{cases}
\label{eq:potdyn}
\end{equation}
where $N_T$ is the total number of electrons in $T$ (i.e., $N_T=10$ for both H$_2$O and NH$_3$) and
we have renamed the constant screening parameter of~(\ref{eq:pot2}) by adding a superscript $c$. 
The idea of this model is that time-dependent screening\footnote{%
We use the terms time-dependent screening and dynamical screening interchangeably in this 
chapter.}
should not affect single ionization but kick
in only after the $q=1$ threshold
has been overcome. It then changes linearly with the degree of ionization until all $N_T$ target electrons
are removed, at which point $N_{\rm X}(P^{\rm{rem}}_{\rm{net}} = N_T)=0$ for all X and
the total potential reduces to a pure
multicenter Coulomb potential [cf.~(\ref{eq:pot2}) and (\ref{eq:pot1})].
The model is similar in spirit to the {\it atomic} dynamical screening model proposed in~\cite{khl+00}.
We explain in Sect.~\ref{sec:pkl} how $P^{\rm{rem}}_{\rm{net}}$, the only time-dependent ingredient, is 
calculated in the present CTMC-IEM approach. 

One can take the dynamical screening 
model to the next level by also including time-dependent
charge screening effects on the projectile ion. Here the rationale is that the screening of the
projectile nucleus increases with an increasing number of captured electrons and hence the
strength of the projectile potential should be down-regulated accordingly. This can be accomplished
by similar modelling as described above with the (fractional) net electron capture $P^{\rm{cap}}_{\rm{net}}$ 
taking the role of $P^{\rm{rem}}_{\rm{net}}$. We have reported on the implementation of 
such a model 
in~\cite{alba_2020}. For the collision systems studied, the effects of dynamical projectile screening
were found to be relatively small 
compared to those of the dynamical
target screening, except in situations in which  
capture into multiply-charged ions is the dominating electron removal
process (see Sect.~\ref{sec:water-tcs}).
We skip the technical details here and refer the interested reader
to~\cite{alba_2020}.

\subsection{Analysis of electron capture and ionization processes}
\label{sec:analysis}
The ingredients of an IEM-based final-state analysis are
single-particle probabilities for the processes of interest and multinomial statistics
to combine them. 
We are interested
in electron capture (cap) and ionization (ion) 
and in the latter case in total and
differential probabilities and cross sections.
The total
probabilities for the $j$th MO ($j=1,\ldots , m$) 
are calculated as the ratios of the number of trajectories contributing to
process $i=$cap or $i=$ion 
to the total number of trajectories
\begin{equation}
	p_j^{\, i} = \frac{n_j^{\, i}}{n_{j,{\rm tot}}} .
\label{eq:psingle}	
\end{equation}
The differential probabilities are calculated by
binning the ionized electrons in small intervals with respect to emission energy 
$\Delta E_{\rm el}$ and emission angle 
$\Delta \Omega_{\rm el}=2\pi[\cos(\theta_{\rm el_{i+1}})-\cos(\theta_{\rm el_{i}})]$~\cite{alba_2019}. 
To arrive at this equation we have exploited the cylindrical symmetry and integrated over the azimuthal angle.
The polar angles $\theta_{\rm el_i}$ and $\theta_{\rm el_{i+1}}$ define the interval for a given $\theta_{\rm el}$: $\theta_{\rm el_i}\leq \theta_{\rm el}< \theta_{\rm el_{i+1}}$.
In extension of~(\ref{eq:psingle}) the single-particle ionization probability associated with $\Delta E_{\rm el}$ and
$\Delta \Omega_{\rm el}$ is
\begin{equation}
\frac{{\rm{d}}^2p_j^{\rm ion}}{{\rm{d}} E_{\rm{el}}{\rm{d}}\Omega_{\rm{el}}}=
\frac{n_j^{\rm ion}}{n_{j,{\rm tot}} \Delta E_{\rm el} 2\pi[\cos(\theta_{\rm el_{i+1}})-\cos(\theta_{\rm el_{i}})]} .
\label{eq:psingled}
\end{equation}

\subsubsection{Differential $q$-fold ionization}
A situation of interest is the detection of one electron at $\{\Delta E_{\rm el},\Delta \Omega_{\rm el}\}$, 
while a total of $q$ electrons are ionized in the same event. In the simplest case of $q=1$ the multinomial
combination of single-particle probabilities yields 
\begin{equation}
	\frac{{\rm{d}}^2P_{q=1}^{\rm ion}}{{\rm{d}} E_{\rm{el}}{\rm{d}}\Omega_{\rm{el}}} =  2 \sum_{j =1}^m{\frac{{\rm{d}}^2 p_{j}^{\rm ion}}{{\rm{d}} E_{\rm{el}}{\rm{d}}\Omega_{\rm{el}}} (1-p_{j}^{\rm ion})
\prod_{k \ne j}^m {(1-p_k^{\rm ion})^2}} .
\label{eq:pdq1}
\end{equation}
One can read this expression as follows: The differential single-electron probability for ionization from the 
doubly-occupied $j$th MO (\ref{eq:psingled}) is multiplied by a non-ionization probability from the same MO and
by a factor of two to account for the spin degeneracy and ensure that exactly one out of two electrons of that MO end up in the continuum. The
resulting binomial probability is further multiplied by non-ionization probabilities from all other MOs and summed 
over $j$ since the detected continuum electron can originate from any of the initially occupied MOs.
For higher $q$, the expressions follow the same logic but become more complicated. The cases
$q=2$ and $q=3$ are shown explicitly in~\cite{alba_2019}. It is useful to define a {\it net}
probability as the weighted sum of the $q$-fold probabilities
\begin{equation}
	\frac{{\rm{d}}^2P_{\rm net}^{\rm ion}}{{\rm{d}} E_{\rm{el}}{\rm{d}}\Omega_{\rm{el}}} =   
	\sum_{q =1}^{N_T} q \frac{{\rm{d}}^2P_{q}^{\rm ion}}{{\rm{d}} E_{\rm{el}}{\rm{d}}\Omega_{\rm{el}}} .
\label{eq:pdiffnet1}
\end{equation}
The net probability corresponds to a measurement in which one detects one electron at $\{\Delta E_{\rm el},\Delta \Omega_{\rm el}\}$ but does not determine if and how many additional electrons are ionized in the same event.
It can also be calculated directly from the differential single-particle probabilities
\begin{equation}
	\frac{{\rm{d}}^2P_{\rm net}^{\rm ion}}{{\rm{d}} E_{\rm{el}}{\rm{d}}\Omega_{\rm{el}}} =   
	2 \sum_{j=1}^m \frac{{\rm{d}}^2p_{j}^{\rm ion}}{{\rm{d}} E_{\rm{el}}{\rm{d}}\Omega_{\rm{el}}} ,
\label{eq:pdiffnet2}
\end{equation}
which provides a useful consistency check.

\subsubsection{Charge-state correlated and inclusive probabilities}
\label{sec:pkl}
The total probabilities (\ref{eq:psingle}) are calculated at each time
step, concurrently
with the ensembles of trajectories, using the energy
criterion described above. At the final time $t=t_f$ they 
form the basis of the 
analysis of single and multiple ionization and capture events within
the IEM. The most straightforward approach is standard multinomial
statistics for all charge-state correlated channels in which a
number of $k$ electrons are captured and $l$ electrons are
ionized to the continuum. The corresponding transition
probabilities $P_{kl}$ (which are functions of the impact parameter vector)
can be written as
\begin{equation}
\begin{alignedat}{1}
P_{kl}=&\sum_{k_1,...,k_m=0}^{M_1,...,M_m} \sum_{l_1,...,l_m=0}^{M_1,...,M_m}\delta_{k,\sum_i k_i}\delta_{l,\sum_i l_i}\prod_{i=1}^{m}{M_i\choose k_i+l_i}{k_i+l_i \choose k_i}\\
&(p_i^{\rm{cap}})^{k_i}(p_i^{\rm{ion}})^{l_i}(1-p_i^{\rm{cap}}-p_i^{\rm{ion}})^{M_i-k_i-l_i},
\end{alignedat}
\label{eq:pkl}
\end{equation}
where $\delta_{k,\alpha}$ is the Kronecker delta symbol and $M_1=M_2=...=M_m=2$ refer to the number of electrons in each MO. 
One problem of this analysis is that the $N_T$ target electrons are distributed over the regions of 
space purely statistically, regardless of whether the projectile can accommodate a given number of
electrons or not. This problem is most prominent for proton projectiles ($Z_p=1$), which can 
capture one or two electrons, but not more than that. In fact, double capture leading to
the formation of a negatively-charged hydrogen ion is already an
extremely rare process which is known to involve electron correlations and which cannot
be described at the level of the IEM, i.e., the multinomial evaluation of $P_{kl}$ 
fails for $k\ge 2$ in the case of proton projectiles. One can try to correct for this by combining the single-particle probabilities
in alternative ways such that the unphysical capture channels are 
closed~\cite{khl+00, Murakami_2012a, alba_2020}, 
but such approaches can be criticized for their ad hoc character and they may 
show deficiencies in
some of the open channels as a consequence of 
re-distributing the total capture flux among the physically allowed $k,l$ combinations. 
In other words,
there is no easy fix to the problem of unphysical multiple capture 
in the IEM and we will not consider it further
in this contribution.

The more inclusive $k$-fold capture probabilities $P_k^{\rm{cap}}$ are given as
\begin{equation} 
P_k^{\rm{cap}}=\sum_{i=0}^{N_T-k}P_{ki} ,
\end{equation}
and can be used to calculate the net capture $P_{\rm{net}}^{\rm{cap}}$ 
\begin{equation} 
P_{\rm{net}}^{\rm{cap}}=\sum_{k=1}^{N_T}k P_{k}^{\rm{cap}} .
\end{equation}
This is the analogue of~(\ref{eq:pdiffnet1}) for differential ionization. The analogue of~(\ref{eq:pdiffnet2}) is
\begin{equation} 
P_{\rm{net}}^{\rm{cap}}=2 \sum_{j=1}^{m} p_{j}^{\rm{cap}}  ,
\label{eq:netcap}
\end{equation}
and similar equations for the total $l$-fold and net ionization probabilities can be
established as well. We use~(\ref{eq:netcap}) and its ionization counterpart to
calculate the time-dependent net capture $P_{\rm{net}}^{\rm{cap}}$, 
net ionization $P_{\rm{net}}^{\rm{ion}}$, and net removal
$P_{\rm{net}}^{\rm{rem}}=P_{\rm{net}}^{\rm{cap}} + P_{\rm{net}}^{\rm{ion}}$, which feed into
the dynamical screening models described in Sect.~\ref{sec:dynamic}.

As a final comment we note that within the semiclassical impact-parameter approximation
differential and total cross sections are obtained by
integrating the probabilities for the processes of interest over the impact parameter vector.

\section{Results and discussion}
\label{sec:results}
In this section, we present a selection of the results obtained from the
CTMC mean-field approach described above. Our goal is to provide the
reader with an overview and a general idea of what has been
and can be accomplished in this framework, and where gaps in our
understanding remain. The results are taken from our original
research papers~\cite{alba_2019, alba_2020, hjl_2020, alba_2023} 
to which we refer the reader for more detailed discussions.

\subsection{Collisions with water molecules: differential cross sections}
\label{sec:water-dcs}
We begin with differential electron emission
from water molecules impacted by highly-charged high-velocity
projectile ions. A useful parameter for this discussion is the
ratio of the projectile charge to the projectile speed 
$\eta = Z_p/v_p$ measured in atomic units.
Low values of $\eta$ correspond to weak perturbations
associated with little ionization of the target molecule. Electron capture
is even less likely and can be ignored in the situations studied 
in this subsection. With increasing $\eta$
one finds an increasing net ionization cross section, a growing fraction
of which is due to multiple ionization events. Accordingly, we can expect
time-dependent screening effects to increase in importance with $\eta$. 
The four collision systems studied here comprise the range $0.47 \le \eta \le 1.03$.
They are 4 MeV/amu C$^{6+}$ ($\eta =0.47$), 3.75 and 3.0 MeV/amu O$^{8+}$
($\eta=0.65$ and 0.73), and 4 MeV/amu Si$^{13+}$ ($\eta=1.03$) projectiles
and were studied experimentally in~\cite{bbb+16,bbm+17,bbc+18}. Along with the data, perturbative
quantum-mechanical calculations were reported, and we compare our results
with those as well. The latter comparison is interesting since the two
approaches complement each other: Ours is based on the $\hbar \rightarrow 0$ limit
of quantum mechanics and uses a three-center target potential; the approach 
of~\cite{bbb+16,bbm+17,bbc+18} uses 
the quantum-mechanical continuum distorted-wave with eikonal initial-state (CDW-EIS)
approximation coupled with a model in which molecular cross sections are given as
linear combinations of atomic contributions. 
We note that a recently reported multicenter CDW-EIS model overcomes some of the 
limitations of the earlier distorted-wave calculations, but the first results are
somewhat inconclusive~\cite{Gulyas23}.
Just like our approach, all these CDW-EIS models are based on single-electron Hamiltonians,
i.e., are of IEM type. 

\begin{figure}
\begin{center}
\resizebox{0.8\textwidth}{!}{\includegraphics{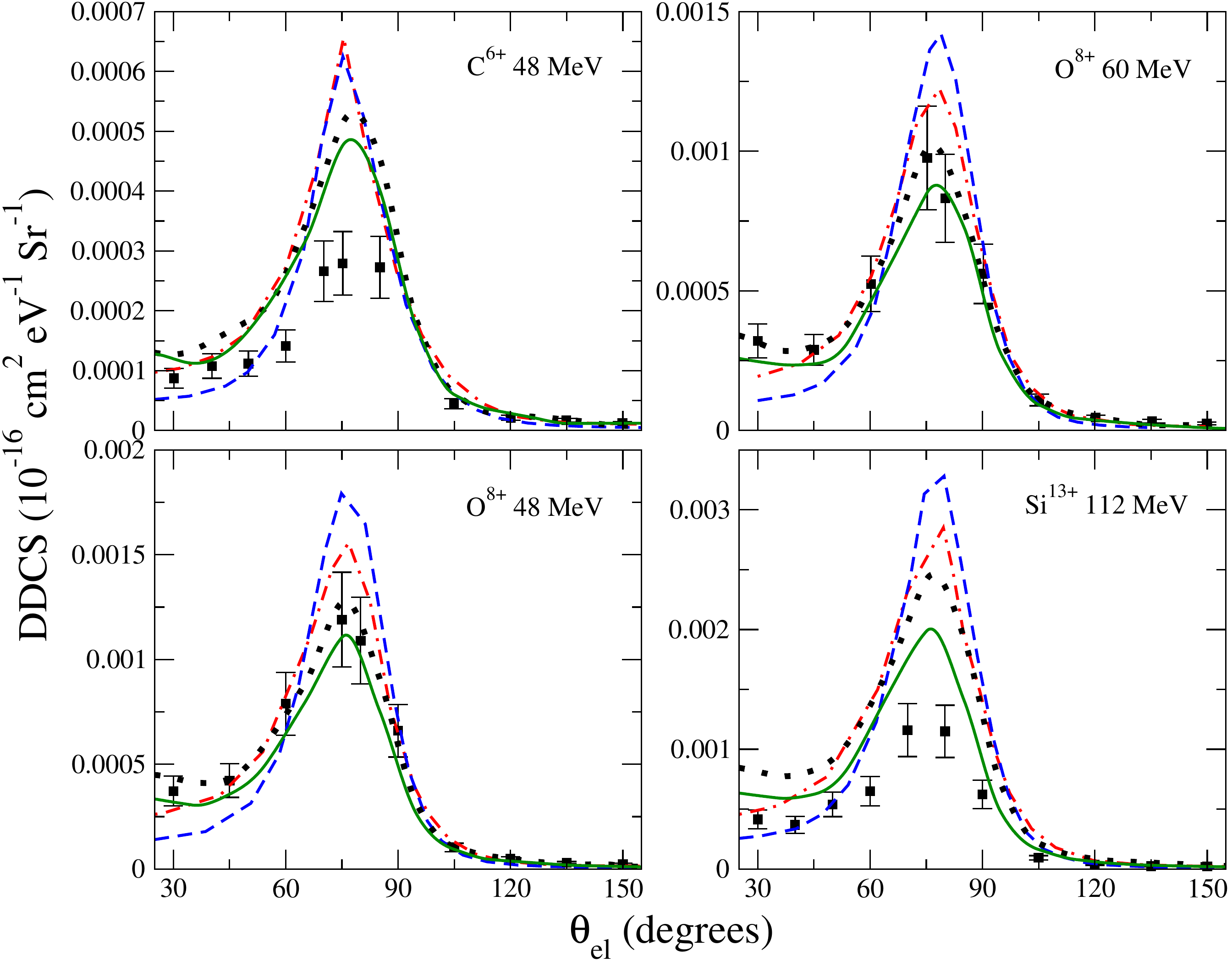}}
\vskip -0.5 truecm
\caption{%
DDCS for net ionization of water molecules as a function of electron emission angle in degrees, 
for the electron emission energy $E_{\rm{el}}=$ 200 eV. The CTMC results 
(obtained with angular resolution $\Delta \theta_{\rm{el}}=10^\circ$) are shown 
as dotted black (static screening) and solid green (dynamical screening)
lines. The experimental data are shown as solid squares with error bars, 
and the prior and post CDW-EIS results are given by dashed-dotted red lines and 
dashed blue lines, respectively~\cite{bbb+16,bbm+17,bbc+18}.
Adapted from~\cite{alba_2019}}
\label{fig:fig1}   
\end{center}
\end{figure}

In Fig.~\ref{fig:fig1}, doubly-differential cross sections (DDCSs) for net 
ionization are shown for the four collision systems. The electron emission
energy is fixed at $E_{\rm el}=200$ eV and the DDCSs are shown as functions
of the polar emission angle $\theta_{\rm el}$ [cf.~(\ref{eq:psingled})].
As expected, there is an overall increase of the DDCS with $\eta$ (note that
the $y$-axis scales are different in the four panels). 
The time-dependent screening results are lower than those obtained with
the static model potential, which is a reflection of the fact that
the target potential becomes more attractive with decreasing charge screening
parameters. At backward angles $\theta_{\rm el} >  90^\circ$ the gap between both
sets of results appears to widen with increasing $\eta$.

Truly quantum-mechanical effects are known to be the strongest at
low and less important at higher electron emission energies, 
and indeed we find good overall agreement
between the CTMC calculations
and the experimental data at $E_{\rm el}=200$ eV. One may deem this energy value as medium
given that it corresponds to a velocity of 3.8 a.u., to be contrasted with
projectile speeds between 11.0 and 12.6 a.u. 
However, the level of agreement in the sequence of the four systems is not as expected:
It is almost perfect for the O$^{8+}$ projectiles, i.e., for the
two intermediate $\eta$ values, but the CTMC data appear to overestimate the
measurements for the other two cases, C$^{6+}$ and Si$^{13+}$ impact at 4 MeV/amu,
regardless of whether the static or time-dependent screening models are used.
The overestimation appears to be independent of the emission angle, which might
indicate that it is due to a problem with the overall normalization of the
experimental data of~\cite{bbm+17}. 

Regarding the comparison with the CDW-EIS calculations, we note that two versions
of the latter are shown, post and prior, which differ in the way the interaction between
the ionized electron and the residual target ion is taken into account. In the post version,
this interaction is purely Coulombic (using an effective charge), while in the
prior form the influence of the non-ionized electrons on the dynamics of the continuum
electron is included more explicitly. Hence, from a formal perspective
the prior form is deemed superior~\cite{bbm+17}.

As a general trend across all four systems we find the CTMC results at small (forward) angles 
to be slightly higher and in the peak region (around $\theta_{\rm el}=60-90^\circ $) 
lower than both CDW-EIS data sets, while there are no significant differences at backward angles.
This gives the CTMC results a slight edge, 
since they match the angular shape of the experimental data very well in
all cases.
The CDW-EIS model for ion-atom collisions is known to be very good at accounting for 
two-center effects, i.e., the combined role of both the projectile and the residual
target ion in shaping the angular distribution of medium-energy 
electrons~\cite{Fainstein_1991, sdr97}.
The present CTMC approach considers projectile and target interactions on
the same footing as well, but in addition takes the three-center geometry of the
target molecule fully into account. We are thus led to conclude that the angular
distributions mirror a {\it many-center} electron-emission mechanism.
Of the available calculations at $E_{\rm el}=200$ eV
only the present CTMC approach is able to describe this properly. 

\begin{figure}
\begin{center}
\resizebox{0.8\textwidth}{!}{\includegraphics{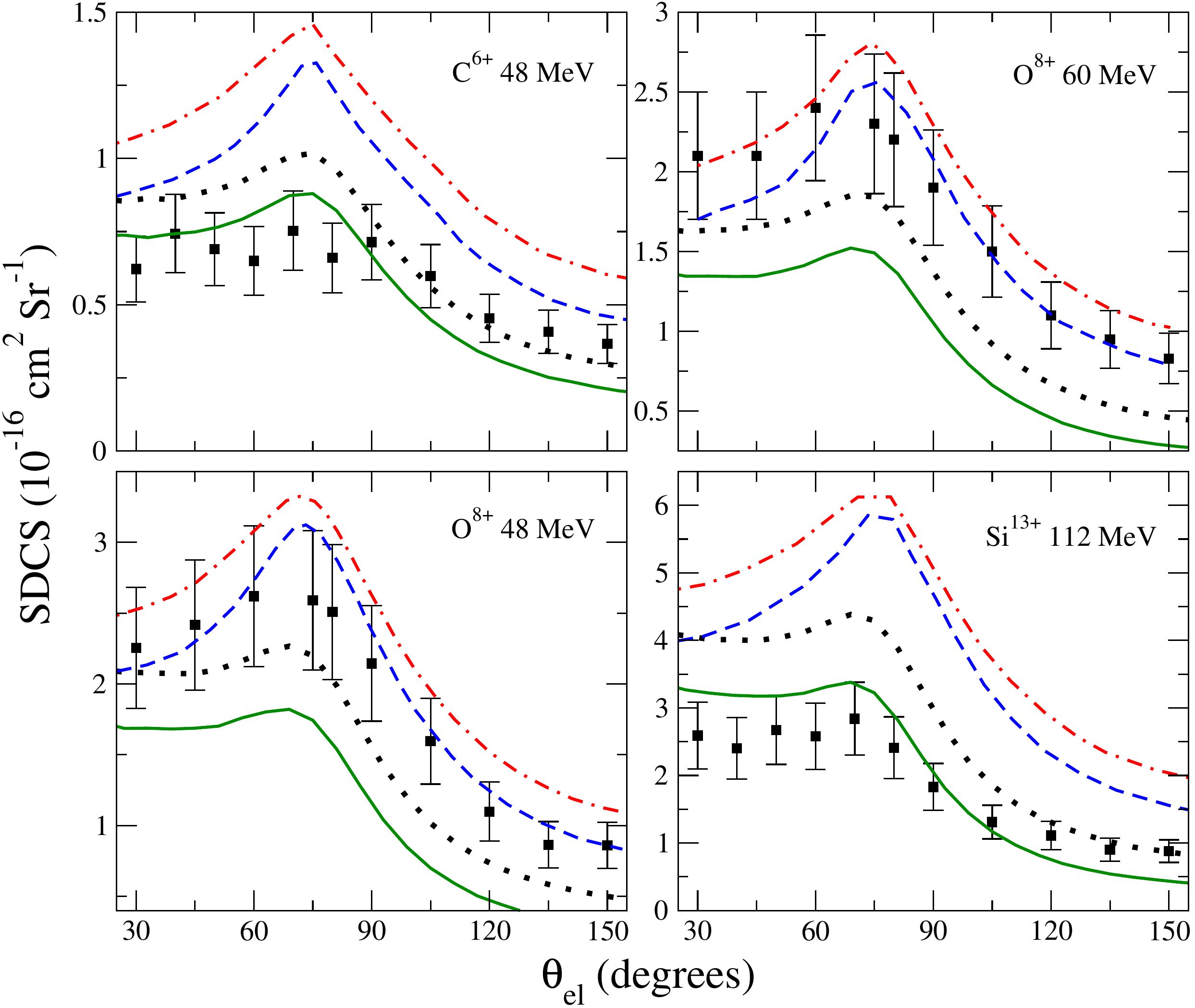}}
\vskip -0.5 truecm
\caption{%
SDCS for net ionization of water molecules as a function of electron emission angle in degrees, 
for the electron emission energy $E_{\rm{el}}=$ 200 eV. The CTMC results 
(obtained with angular resolution $\Delta \theta_{\rm{el}}=6^\circ$) are shown 
as dotted black (static screening) and solid green (dynamical screening)
lines. The experimental data are shown as solid squares with error bars, 
and the prior and post CDW-EIS results are given by dash-dotted red lines and 
dashed blue lines, respectively~\cite{bbb+16,bbm+17,bbc+18}.
Adapted from~\cite{alba_2019}}
\label{fig:fig2}   
\end{center}
\end{figure}

The same conclusion can be drawn from Fig.~\ref{fig:fig2}, which displays
singly-differential cross sections (SDCSs) as functions of the emission angle
for the four collision systems. 
A closer analysis of the angular distributions in terms of a comparison of
ratios of cross sections at forward vs intermediate scattering angles reveals that
the CTMC results give a better representation of the experimental data than 
CDW-EIS, the prior version of which fares better than the
post form, as expected~\cite{alba_2019}. 

On a quantitative level, there appears to be an inconsistency in the SDCSs of
Fig.~\ref{fig:fig2} and the DDCSs of Fig.~\ref{fig:fig1} regarding the 
comparison between the CTMC and the experimental results:
For the O$^{8+}$ projectiles we found very good agreement for the DDCSs
at $E_{\rm el}=200$ eV, but the SDCSs are underestimated. Likewise, for the
other two projectiles the CTMC data appear to overestimate the
measured DDCSs, but they are relatively close to the experimental SDCSs. 
These observations are, however, quite consistent, since the CTMC approach
is known to fall short for low-energy electron emission. This is discussed
in more detail in~\cite{alba_2019} in which doubly-differential
results at $E_{\rm el}=5$ eV are shown in addition to the $E_{\rm el}=200$ eV
data. One may conclude that an underestimation of the experimental SDCS is to be expected,
and the fact that we do not observe it for C$^{6+}$ and Si$^{13+}$ projectiles
points again to a possible overall normalization problem of the experimental data.

\begin{figure}
\begin{center}
\resizebox{0.8\textwidth}{!}{\includegraphics{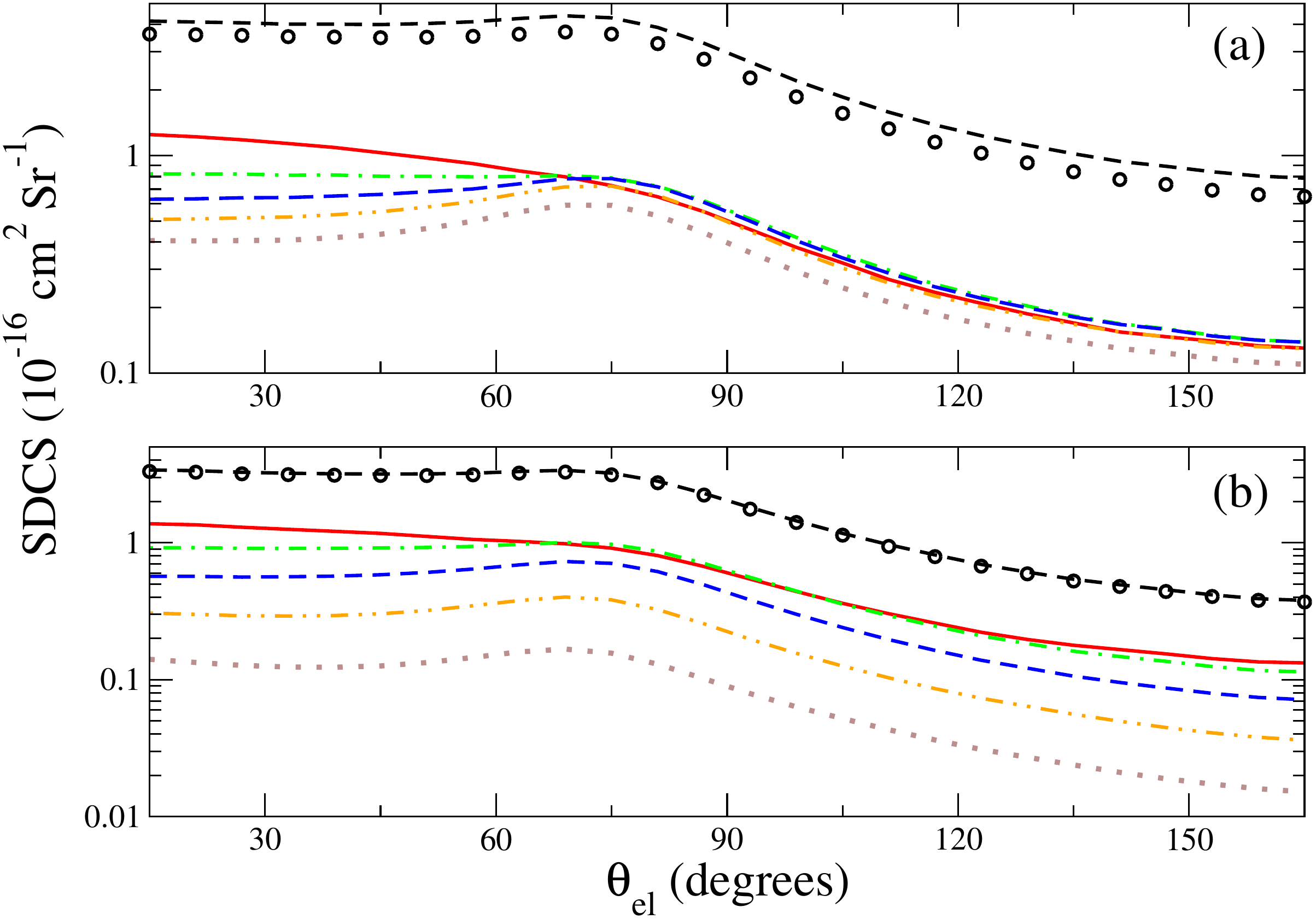}}
\vskip -0.5 truecm
\caption{%
SDCS for net ionization of water molecules as a function of electron emission angle in degrees
for 112~MeV Si$^{13+}$ projectiles with (a) static and (b) dynamical screening.
($---$) Net, ({\textcolor{red}{---}}) $\frac{{\rm{d}} \sigma_1}{{\rm{d}} \Omega_{\rm{el}}}$, ({\textcolor{green}{$\cdot - \cdot$}}) $2\frac{{\rm{d}} \sigma_2}{{\rm{d}} \Omega_{\rm{el}}}$, ({\textcolor{blue}{$---$}}) $3\frac{{\rm{d}} \sigma_3}{{\rm{d}} \Omega_{\rm{el}}}$, ({\textcolor{orange}{$\cdot\cdot -\cdot\cdot$}}) $4\frac{{\rm{d}} \sigma_4}{{\rm{d}} \Omega_{\rm{el}}}$, ({\textcolor{brown}{$\cdot\cdot\cdot$}}) $5\frac{{\rm{d}} \sigma_5}{{\rm{d}} \Omega_{\rm{el}}}$, ($\circ$) $\sum\limits_{q=1}^5q\frac{{\rm{d}} \sigma_q}{{\rm{d}} \Omega_{\rm{el}}}$.
Adapted from~\cite{alba_2019}}
\label{fig:fig3}   
\end{center}
\end{figure}

As for the static vs time-dependent screening model comparison, we once again
observe that the latter results in less ionization than the former
and that this effect increases with $\eta$. 
In order to shed more light on the role of time-dependent screening 
we show in Fig.~\ref{fig:fig3} for both CTMC models the contributions from the first five
$q \frac{d\sigma_q}{d\Omega_{\rm el}}$ terms to the net SDCS [cf.~(\ref{eq:pdiffnet1})]
as well as their
sum and the full net cross section
for the Si$^{13+}$ projectile for which $\eta$ assumes the largest value
among the four collision systems studied ($\eta =1.03$).

For the $\sigma_1$ and $2\sigma_2$ terms the static and time-dependent
screening results are similar, but higher-order contributions are suppressed
in the time-dependent screening model and increasingly so with increasing
electron multiplicity $q$.
As a result, the sum truncated at $q=5$ is very
close to the full net cross section, i.e., contributions from $q>5$ account for 
at most 3\% in the time-dependent screening model. 
For the static case those contributions range between 13\% at
small emission angles and 18\% at backward angles.  

\begin{figure}
\begin{center}
\resizebox{0.8\textwidth}{!}{\includegraphics{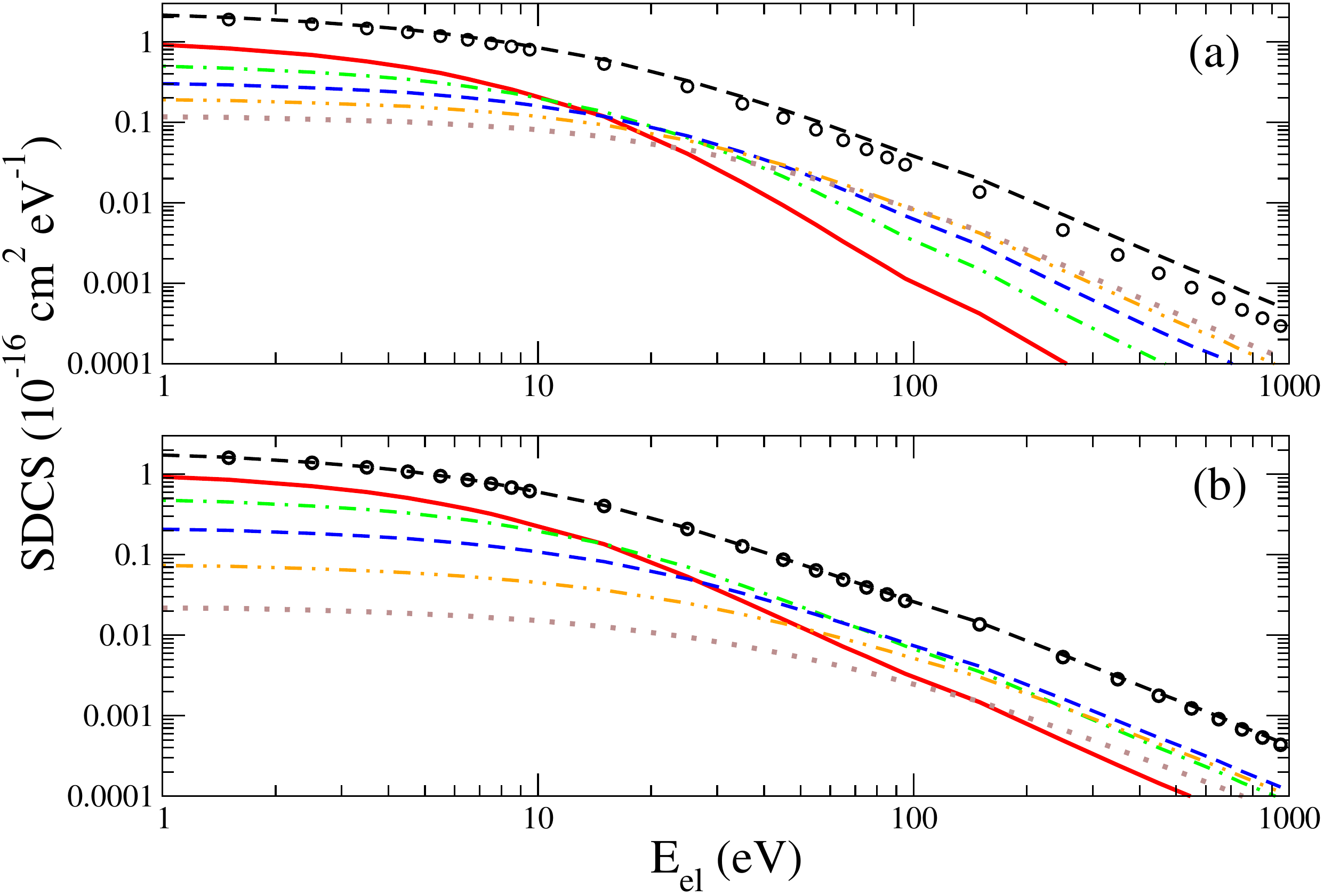}}
\vskip -0.5 truecm
\caption{%
SDCS for net ionization of water molecules as a function of electron emission energy
for 112~MeV Si$^{13+}$ projectiles with (a) static and (b) dynamical screening.
($---$) Net, ({\textcolor{red}{---}}) $\frac{{\rm{d}} \sigma_1}{{\rm{d}} E_{\rm{el}}}$, ({\textcolor{green}{$\cdot - \cdot$}}) $2\frac{{\rm{d}} \sigma_2}{{\rm{d}} E_{\rm{el}}}$, ({\textcolor{blue}{$---$}}) $3\frac{{\rm{d}} \sigma_3}{{\rm{d}} E_{\rm{el}}}$, ({\textcolor{orange}{$\cdot\cdot -\cdot\cdot$}}) $4\frac{{\rm{d}} \sigma_4}{{\rm{d}} E_{\rm{el}}}$, ({\textcolor{brown}{$\cdot\cdot\cdot$}}) $5\frac{{\rm{d}} \sigma_5}{{\rm{d}} E_{\rm{el}}}$, ($\circ$) $\sum\limits_{q=1}^5q\frac{{\rm{d}} \sigma_q}{{\rm{d}} E_{\rm{el}}}$.
Adapted from~\cite{alba_2019}}
\label{fig:fig4}   
\end{center}
\end{figure}

A complementary view is provided by a similar comparison of cross sections
differential in
electron emission energy instead of emission angle. Such a comparison is
shown in Fig.~\ref{fig:fig4}. At low electron energies the contribution
from the excluded $q>5$ terms is very small in both screening models, but 
the situation is different at higher $E_{\rm el}$. At the highest energies 
shown, the contribution from those high-$q$ terms is almost as large as
that from $q\le 5$ in the case of the static model, whereas it stays below
10\% if time-dependent screening is turned on. It would be of great
interest if differential experimental data for fixed $q$ would become
available to test these predictions.

For proton-H$_2$O collisions a comparison of the present CTMC approach was carried 
out with recent measurements and also with CDW-EIS 
theory (including a new model)\cite{Bhogale_2022}. 
This comparison demonstrates that the multicenter CTMC approach works very well when dealing with
mostly single ionization, and complements the multiply charged projectile results discussed in this section.

\subsection{Collisions with water molecules: total cross sections}
\label{sec:water-tcs}
\begin{figure}
\begin{center}
\resizebox{0.8\textwidth}{!}{\includegraphics{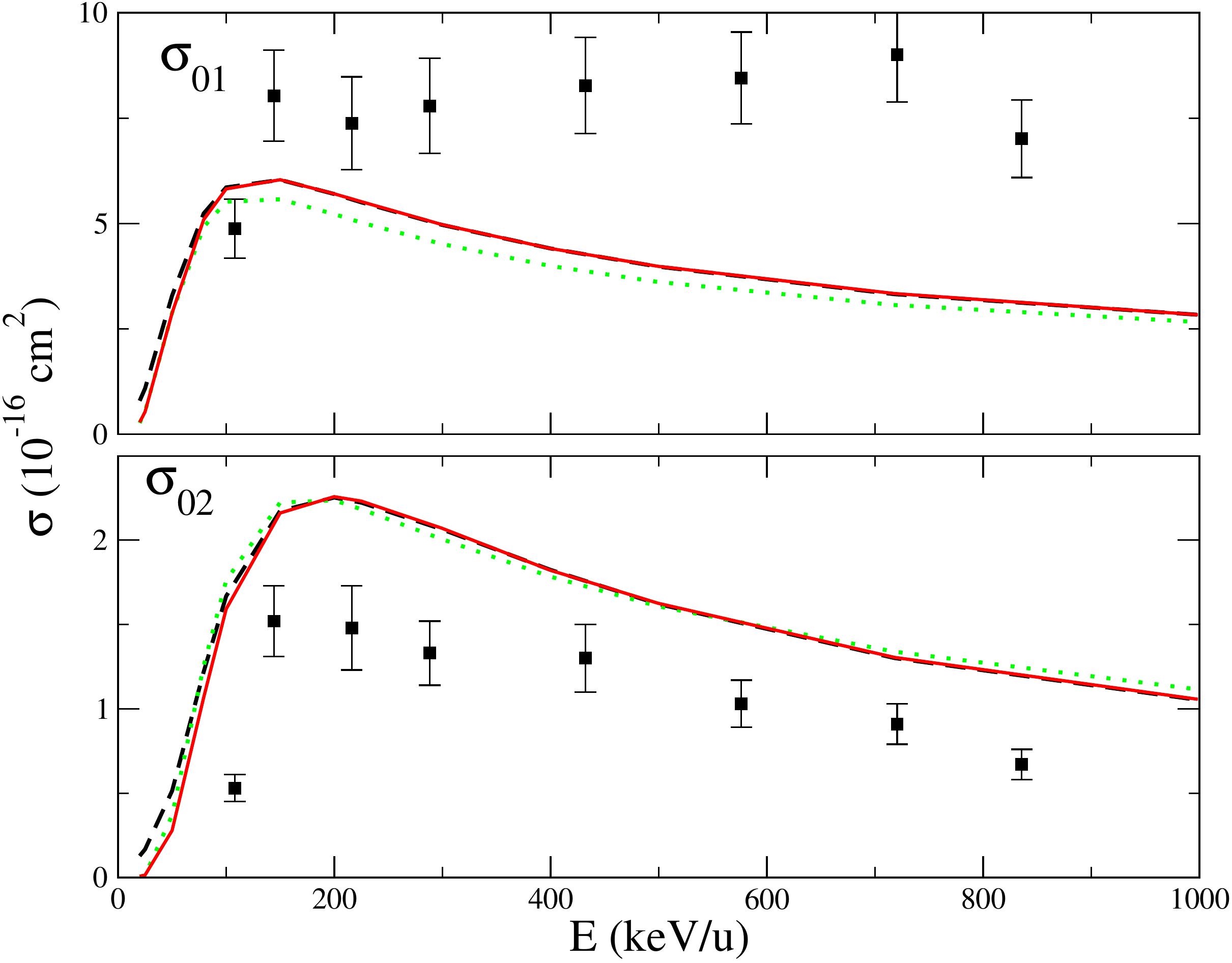}}
\vskip -0.5 truecm
\caption{%
Pure single and double ionization cross sections for Li$^{3+}$-H$_2$O collisions
as functions of impact energy.
The CTMC results are shown 
as dashed black (dynamical target and projectile screening), solid red (dynamical target screening),
and dotted green (static screening) lines. 
The experimental data are shown as solid squares with error bars~\cite{lwm+16}. 
Adapted from~\cite{alba_2020}}
\label{fig:fig5}   
\end{center}
\end{figure}

We now turn our attention to total cross sections for processes
that involve the removal of one or several electrons.
Figure~\ref{fig:fig5} displays the pure ionization cross sections
$\sigma_{01}$ and $\sigma_{02}$ calculated with the trinomial
formula (\ref{eq:pkl}) for the Li$^{3+}$-H$_2$O collision system.
Three CTMC data sets are shown: static screening results,
results obtained from using the time-dependent target charge screening
parameters (\ref{eq:potdyn}), and results from calculations in which
time-dependent projectile screening effects are also included.
Obviously, the three models give very similar results
over the range of impact energies shown 
and they do not agree very well with the experimental data. The underestimation
of $\sigma_{01}$ can, once again, be explained by a known weakness of the
CTMC approach, namely a shortfall at large impact parameters which dominate
singly ionizing collisions. Since ionization at large impact parameters is mostly
associated with
low-energy electron emission, this is the same weakness as the one identified in 
Sect.~\ref{sec:water-dcs} above.

\begin{figure}
\begin{center}
\resizebox{0.8\textwidth}{!}{\includegraphics{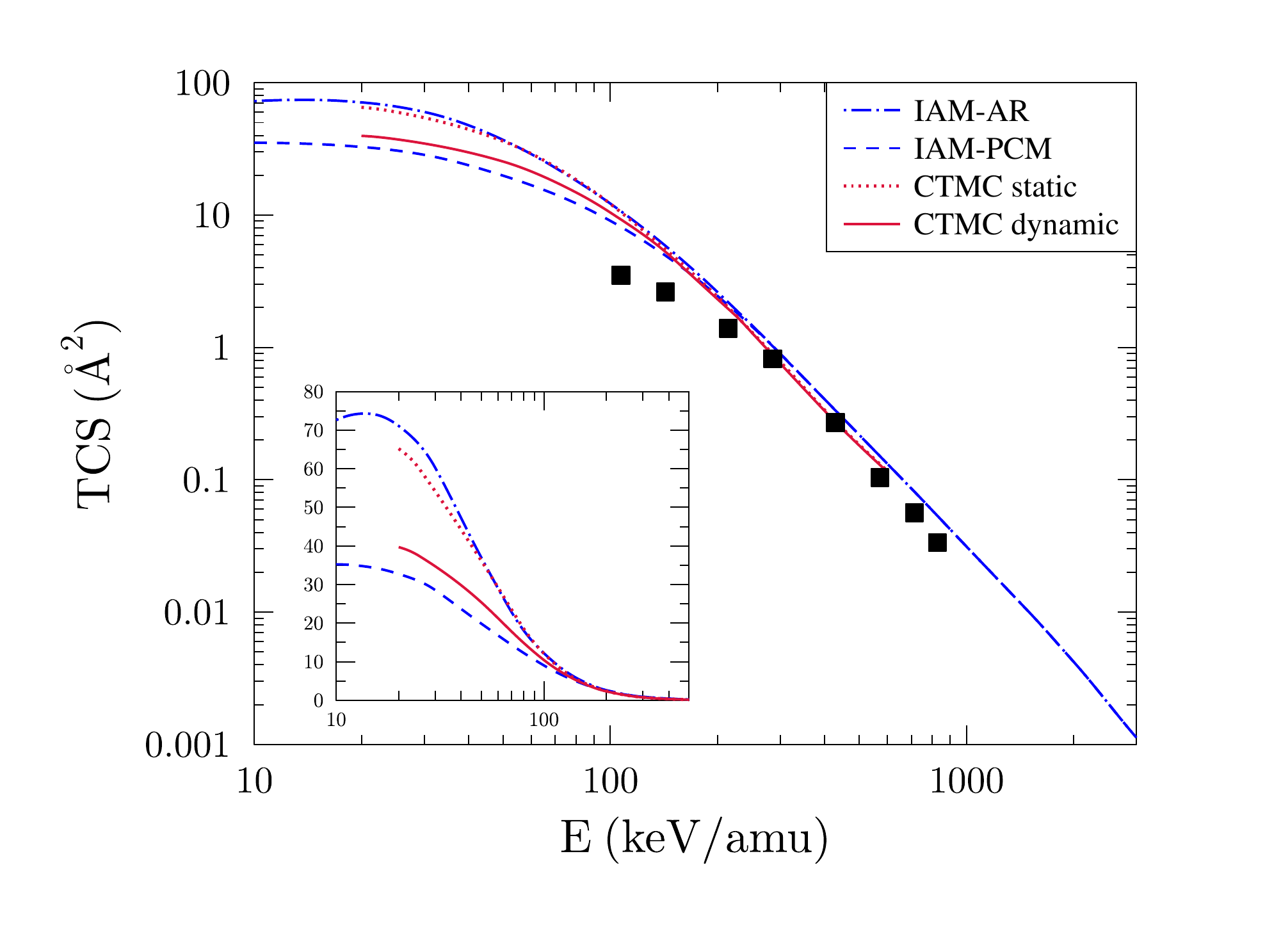}}
\vskip -1.2 truecm
\caption{%
Net electron capture cross section for Li$^{3+}$-H$_2$O collisions as a
function of impact energy.
The highest curve (dash-dotted blue) shows the additivity rule result (IAM-AR), 
the lowest (dashed blue) curve the IAM-PCM result. In between are the CTMC results, 
namely the (dotted red) static screening, and below it 
(solid red) the dynamical screening result.
The experimental data are shown as black solid squares~\cite{lwm+16}.
Adapted from~\cite{hjl_2020}}
\label{fig:fig6}   
\end{center}
\end{figure}

Figure~\ref{fig:fig6} shows the net capture cross section for the
same collision system. Only static screening (`CTMC static') and time-dependent
target {\it and} projectile screening (`CTMC dynamic') results are included, but
we note that the target-only time-dependent screening cross section
is very close to the 'CTMC dynamic' data shown.

At high impact energies, the static and dynamic results are very similar --- there
simply is not enough time for the mean-field potential to adapt and 
thereby change the character of the trajectories.
Below $E\approx 100$ keV/amu, however, 
time-dependent screening effects are strong, as can be seen most clearly in
the inset which displays the cross section on a linear scale.

In addition to the CTMC results the figure
includes experimental data from~\cite{lwm+16} and results from two sets of quantum-mechanical 
independent-atom-model (IAM) calculations: `IAM-AR`, in which the simple
additivity rule is used to sum up atomic cross sections, and `IAM-PCM', 
which is a much more sophisticated model which takes the geometric overlap
of the atomic cross sections into account~\cite{hjl_2016}.
Interestingly, the static-potential CTMC results agree well with the IAM-AR data, 
while
the time-dependent screening calculations are closer to the IAM-PCM cross sections.
A similar pattern was observed for He$^{2+}$ impact~\cite{hjl_2020}.
At $E\ge 200$ keV/amu the results from all four models merge and show good agreement
with the experimental data. 
As discussed in~\cite{hjl_2020}, it is difficult to understand from a modelling perspective
why all calculated cross section curves for Li$^{3+}$ impact are significantly above 
the two experimental data points at $E< 200$ keV/amu. This observation is
also in striking contrast to what was found for He$^{2+}$ impact and warrants
further investigation, ideally aided by additional experimental work.

\begin{figure}
\begin{center}
\resizebox{1.0\textwidth}{!}{\includegraphics{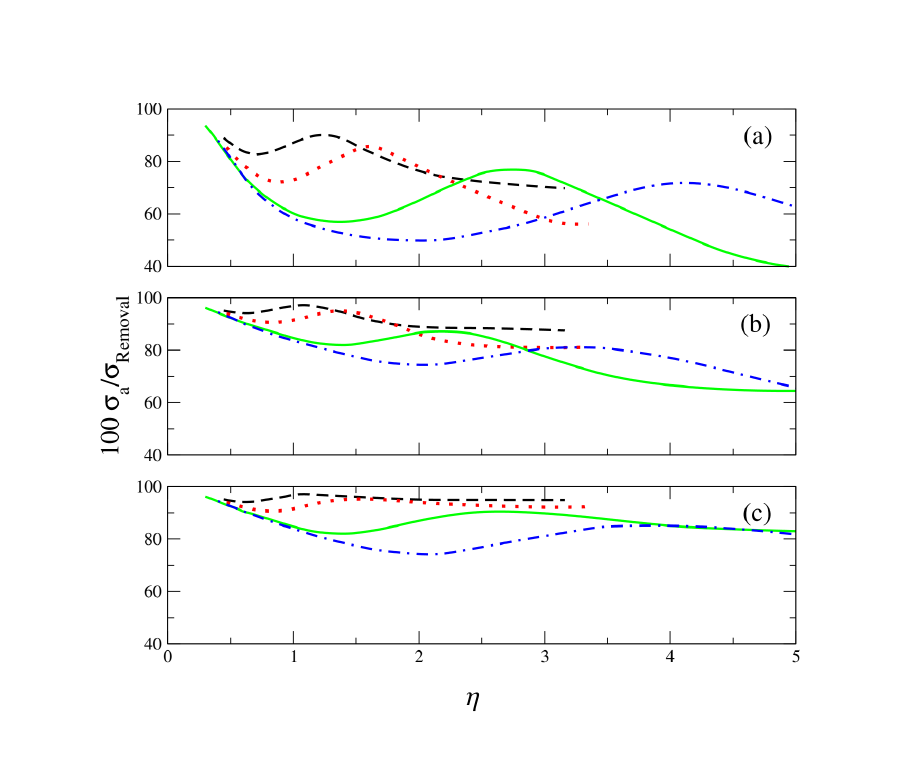}}
\vskip -1.5 truecm
\caption{%
Percentage of $\sum_{q=1}^3 q\sigma_q^{\rm rem} = \sum_{q=1}^3 q(\sigma_q^{\rm cap} + \sigma_q^{\rm ion})$ 
with respect to the total net removal cross section $\sigma_{\rm net}^{\rm rem}$ as a function
of $\eta=Z_p/v_p$
in the case of (a) static screening,
(b) dynamical target screening, (c) dynamical target and projectile screening potentials. 
The systems shown are He$^{2+}$ ($---$),
Li$^{3+}$ ({\textcolor{red}{$\cdot\cdot\cdot$}}),
C$^{6+}$ ({\textcolor{green}{---}}),
Ne$^{10+}$ ({\textcolor{blue}{$\cdot - \cdot$}}).
Adapted from~\cite{alba_2020}}
\label{fig:fig7}   
\end{center}
\end{figure}

We end this subsection with a comparison between static and dynamic screening
model results which uses similar ideas as in
Figs.~\ref{fig:fig3} and \ref{fig:fig4}, but this time on the basis of {\it total}
cross sections:
Figure~\ref{fig:fig7} displays for four different projectiles 
(He$^{2+}$, Li$^{3+}$, C$^{6+}$, Ne$^{10+}$)
the percentage
contribution of $\sigma_a = \sum_{q=1}^3 q \sigma_q^{\rm rem}$ to the total net
removal cross section $\sigma_{\rm net}^{\rm rem}$ as a function of the parameter $\eta=Z_p/v_p$. A value
of close to 100\% signals that the $q\ge 4$ contributions to $\sigma_{\rm net}^{\rm rem}$
are negligibly small. This is the situation observed in the perturbative small-$\eta$ regime.
As $\eta$ increases, all the curves in Fig.~\ref{fig:fig7} show a decreasing trend, 
eventually reaching a first minimum located at an $\eta$ value which is larger,
the larger the projectile charge is. The minimum is followed by a local maximum and
then another decrease. A closer analysis of the data shows that for each projectile
the minimum occurs in a region in which
ionization is the dominant electron removal process. Capture becomes competitive around
the maximum and takes over as the dominant process at even higher $\eta$ values. 

For the static screening results displayed in panel (a) the minima are deep, in particular
for the highly-charged projectiles, indicating that high-multiplicity (i.e., $q>3$-fold) 
ionization is predicted to be very strong. Once time-dependent target screening is
included [panel (b)], high-$q$ processes are suppressed and the minima become shallower. 
Additional inclusion of time-dependent projectile screening [panel (c)] affects the
data mostly at large $\eta$ values, i.e., in the region in which
capture becomes dominant.

There is no apparent scaling with respect to $\eta$:
When moving from one projectile to the next, the positions of the minima and maxima change 
in nontrivial ways. 
This is perhaps not surprising in the nonperturbative $\eta \ge 1$ region, but
based on the present $\hbar=0$ mean-field calculations alone we cannot
draw definitive conclusions.
Ultimately, experimental data would be needed for a better understanding of the
strength of the high-multiplicity electron removal processes as a function of $\eta$.

\subsection{Collisions with ammonia molecules}
\label{sec:ammonia}
For the case of ammonia target molecules we focus on collisions 
with protons for which differential and total cross section data
are available for comparison.
As can be expected for a singly-charged projectile, we found dynamical
screening effects to be rather weak, and thus we discuss results obtained
from the static screening model only. Net differential electron emission
was measured some time ago~\cite{Lynch1976}, while total cross section
data for single and multiple ionization events were reported
more recently~\cite{Wolff2020}. Comparing the latter with calculations requires
some caution since the measurements are based on counts for charged
fragments and do not directly correspond to the charge-state correlated
cross sections for capture and ionization discussed in Sect.~\ref{sec:pkl}.

\begin{figure}
\begin{center}
\resizebox{1.0\textwidth}{!}{\includegraphics{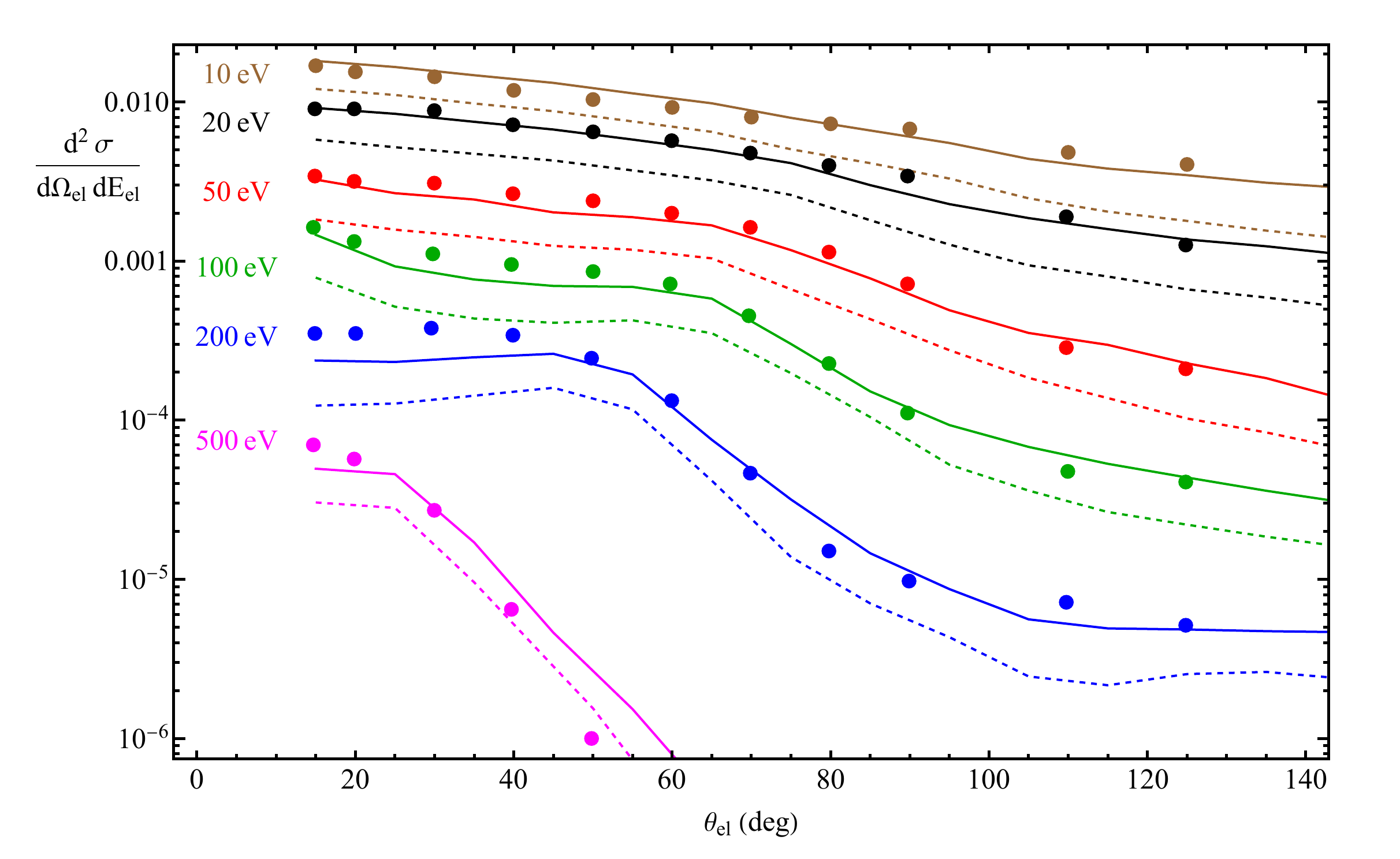}}
\vskip -0.5 truecm
\caption{%
DDCS in units of \AA$^2$/(eV srad) for p-NH$_3$ collisions 
at $E=250$ keV for ionized electron energies of
$E_{\rm el} = 10, 20, 50, 100, 200, 500$ eV. 
Solid lines: CTMC net ionization results obtained with~(\ref{eq:pdiffnet2}); 
dashed lines are for single ionization obtained with (\ref{eq:pdq1}). The data points are the
experimental results of~\cite{Lynch1976}, as reported in~\cite{Senger1988, Mondal2017}. 
Adapted from~\cite{alba_2023}}
\label{fig:fig8}   
\end{center}
\end{figure}

We show in Fig.~\ref{fig:fig8} DDCS results for $E=250$ keV proton impact as
functions of the polar emission angle for a set of electron energies between
$E_{\rm el}=10$ eV and $E_{\rm el}=500$ eV. 
We have calculated both net DDCSs and DDCSs for $q=1$ according to~(\ref{eq:pdiffnet2})
and (\ref{eq:pdq1}) and display them as solid and dashed lines, respectively.
The agreement with the experimental net DDCS results is very good over the entire range of electron angles and energies.
Comparison with previously obtained quantum calculations is made in~\cite{alba_2023}, 
which demonstrates that the present CTMC method which is based on a multicenter effective 
potential provides an improved description at backward electron emission angles. 

The comparison between the solid and dashed lines provides an indication of the
significance of multiple ionization. The net and $q=1$ results are
of similar shape and typically differ by a factor of two, i.e., the calculations
predict that double and higher multiple ionization contribute about equally to
net ionization as the $q=1$ channel. In the absence of $q$-specific DDCS
data we can only speculate whether this is indeed the case in this system
or if the result signals a deficiency of the IEM.
We do note that NH$_3$ is an extended object with a number of 
relatively loosely bound electrons,
i.e., multiple ionization might indeed be quite strong.

\begin{figure}
\begin{center}
\resizebox{0.9\textwidth}{!}{\includegraphics{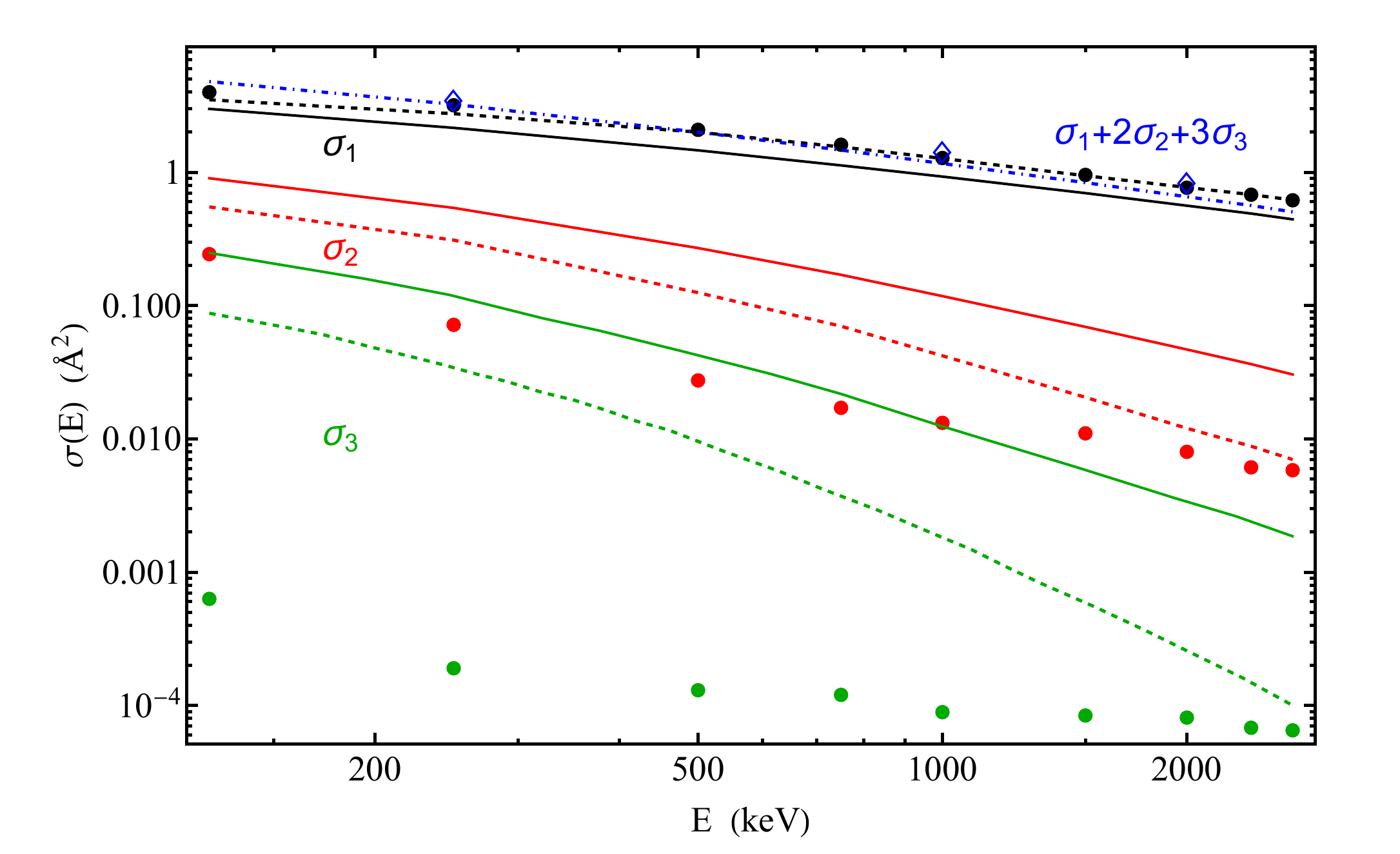}}
\vskip -0.5 truecm
\caption{%
Total cross sections for p-NH$_3$ collisions as
functions of impact energy. 
Dash-dotted blue line: CTMC approximate net ionization result,
$\sum_{q=1}^3 q \sigma_q^{\rm ion}$.
Blue open diamonds: net ionization cross sections from~\cite{Lynch1976}. Solid
lines (black, red, green) from top to bottom: CTMC results for $\sigma_q^{\rm ion}$ with $q = 1, 2, 3$ respectively. Dashed lines
(black, red, green): IAM-PCM results shown in~\cite{Wolff2020} and~\cite{hjl_2022} for $\sigma_q^{\rm ion}$. 
Dots (black, red, green):
experimental values for $\sigma_q^{\rm ion}$ from fragmentation yields~\cite{Wolff2020}. 
The results for $q = 3$ are associated solely with H$^+$ + N$^{2+}$ coincidences.
Adapted from~\cite{alba_2023}}
\label{fig:fig9}   
\end{center}
\end{figure}

We end this section with a look at the total cross sections for net and $q=1,2,3$-fold
ionization in Fig.~\ref{fig:fig9}. The present net results are calculated according to
$\sigma_{\rm net}^{\rm ion} = \sigma_1^{\rm ion} + 2\sigma_2^{\rm ion} +3 \sigma_3^{\rm ion}$,
i.e., contributions from $q\ge 4$ events are neglected. 
IAM-PCM results from~\cite{hjl_2022} are also included in the figure, 
and so are net measurements from~\cite{Lynch1976}
and $\sigma_q^{\rm ion}$ data from~\cite{Wolff2020} obtained by summing up fragmentation yields.

The present net cross section is quite close to the data point at $E=250$ keV, but falls
below the measurements at higher energies. The calculation for $q=1$ produces a
cross section curve which matches the shape of the experimental data of~\cite{Wolff2020} quite
well, but underestimates them by about 30~\%. This is different from the IAM-PCM
prediction which agrees well with the $q=1$ data at intermediate
and high energies and is marginally
below the two data points at $E=125$ keV and $E=250$ keV. Also for $q=2$ and $q=3$ do the
two theories differ---both in shape and in magnitude.

From a comparison of the experimental net data 
of~\cite{Lynch1976} and the $q=1$ data of~\cite{Wolff2020} one would conclude that
the net cross section is
almost entirely due to single ionization, and indeed the experimental
$q=2$ and $q=3$ data included in
Fig.~\ref{fig:fig9} appear to confirm this. However, the fragment measurements 
of~\cite{Wolff2020} may miss some contributing channels, since certain coincidences cannot
be detected with the set-up used. This is to say that
the reasons for the apparent very strong overestimation of double and triple
ionization by the present CTMC as well as the previous IAM-PCM calculations
remain somewhat unclear. We refer the reader to~\cite{alba_2023} for a more detailed discussion
of this issue.

\section{Concluding remarks}
\label{sec:conclusions}
We have presented a mean-field model applied at the $\hbar = 0$ level to deal
with the problem of capture and ionization in ion-molecule collisions.
Both differential and total cross sections were calculated, and a select set of
results for various ions impinging on water and ammonia molecules have been
discussed in comparison with experimental data and some of the
existing previous calculations.
These comparisons shed light on the strengths and weaknesses
of the model and on open questions.

To summarize our findings, we start with noting that the use of multicenter
potentials to describe the molecular targets appears to be a plus compared to 
previous calculations in which the molecular geometry was modelled
in less sophisticated fashions. This is most obvious in angular
electron distributions. The comparisons provided suggest that the notion of two-center
electron emission, extensively discussed for ion-atom collision problems, has to
be replaced by many-center electron emission for molecular targets.

The IEM framework used includes the straightforward calculation of multiple
capture and ionization contributions via multinomial statistics. We have
found that contributions from multiplicities higher than $q=1$ can be quite
substantial, even for singly-charged ion impact. 
To which extent this reflects
the physics at play or a shortcoming of the IEM remains to be seen, given
that the existing coincident experiments for molecular targets are based
on the counting of charged fragments and may suffer from incomplete
detection, i.e., certain contributions might
be missed and others misidentified. Further experimental investigation and
also theoretical efforts which do not rely on the IEM are needed
to shed more light on this problem.

One limitation of the present CTMC model is its inability to describe 
correctly the forward emission of low-energy electrons. This problem
cannot be fixed easily, certainly not without going beyond the
$\hbar = 0$ level of the theory. This, however, appears to be challenging
to say the least, unless one is willing to compromise on other aspects
of the problem at hand, such as its multicenter nature.
It is thus safe to conclude that CTMC models, despite their
imperfections, will continue to play
an important role in aiding the understanding 
of ion-molecule collision problems.

\begin{acknowledgments}
Financial support from the Natural Sciences and Engineering Research Council of Canada (NSERC) 
(RGPIN-2017-05655 and RGPIN-2019-06305) is gratefully acknowledged. 
\end{acknowledgments}

\bibliography{springer}

\end{document}